\newif\ifarxiv
\arxivtrue 

\ifarxiv
	\documentclass[12pt]{article}
	\usepackage[margin=1in]{geometry}	
	\usepackage[slantedGreek]{mathpazo}
	\usepackage[pdftex]{graphicx}
	\usepackage{amsfonts}
	\usepackage{amssymb}
	\usepackage{amsthm}
	\usepackage{float}
	\usepackage{amsmath}%
	\usepackage{hyperref}
	\usepackage{natbib}
	\usepackage{dsfont}
\else
\fi
\usepackage{algorithmic,algorithm}

\newcommand{\nc}{\newcommand}
\nc{\defeq}{\mathrel{\mathop:}=}
\usepackage{color,soul,marginnote}

\newcommand\independent{\protect\mathpalette{\protect\independenT}{\perp}}
\def\independenT#1#2{\mathrel{\rlap{$#1#2$}\mkern2mu{#1#2}}}

\ifarxiv
\title{Generative Bayesian Computation for Maximum Expected Utility}
\author{	
	\makebox[.4\linewidth]{Nick Polson}\\\textit{Booth School of Business}\\\textit{University of Chicago}\\\and 
	\makebox[.4\linewidth]{Fabrizio Ruggeri}\\\textit{Italian National Research}\\\textit{Council in Milano}\\\and
	\makebox[.4\linewidth]{Vadim Sokolov\footnote{Nick Polson is  Professor of Econometrics and Statistics at Chicago Booth: ngp@chicagobooth.edu. Fabrizio Ruggeri is Professor of Statistics at CNR IMATI, Milano, I-20133, Italy. Vadim Sokolov is Associate Professor at Volgenau School of Engineering
George Mason University. vsokolov@gmu.org.} }\\\textit{Department of Systems Engineering and Operations Research}\\\textit{George Mason University}
}
\date{First Draft July 12, 2023 \\This Draft: August 18, 2024}
\else
\fi
\graphicspath{{./fig/}}

\begin{document}

\maketitle
\begin{abstract}
\noindent Generative Bayesian Computation (GBC)  methods are developed to provide an efficient computational solution for  maximum expected utility (MEU).
We propose  a density-free generative method based on quantiles that naturally calculates 
expected utility  as a marginal of  quantiles. 
 Our approach uses a deep quantile neural estimator  to directly estimate distributional utilities.
Generative methods assume only the ability to simulate from the model and parameters and as such are
likelihood-free.  A large training dataset is generated from parameters and output together with  a base distribution.
Our method  a number of 
computational advantages primarily being density-free with an efficient estimator of expected utility.
A link with the  dual theory of expected utility  and risk taking 
is also discussed.. To illustrate our methodology, we solve an optimal portfolio allocation problem with Bayesian learning and 
a power utility (a.k.a. fractional Kelly criterion). Finally, we conclude with directions for future research.
\end{abstract}

\newpage
\section{Introduction}

Generative Bayesian Computation (GBC) constructs a probabilistic map to represent a posterior distribution and to calculate functionals of interest. Our goal here is to extend generative methods  to solve maximum expected utility (MEU) problems. 
We propose a density-free generative method that has the advantage of being able to compute expected utility as a by-product.
To do this, we find a deep quantile neural map to represent the distributional utility. Then we provide a key identity which represents the expected utility as a marginal of quantiles.

Although deep learning has been widely used in engineering \citep{dixon2019deep} and econometrics applications \citep{heaton2017deep} and were shown to outperform classical methods for prediction \citep{sokolov2017discussion} solving optimal decision problems has received less attention.   Our work builds on the reinforcement learning literature \cite{dabney2017,dabney2018} where it is not necessary to know the utilities rather one needs a panel of known rewards and input parameters. The main difference then is our  assumption of a utility function \citep{lindley1976} and its use in architecture design at the first level of the hierarchy.
Recent work on generative methods includes \cite{zammit-mangion2024,sainsbury-dale2024} in spatial settings, \cite{nareklishvili2023generative} for causal modeling
and \cite{polson2024generative} for engineering problems.

Our work also builds on \cite{muller1995} who use curve fitting techniques to solve MEU problems. Our work is also related to the reinforcement learning literature of \cite{dabney2017}.  It differs in that we assume a given utility function, and we directly simulate and model the random utilities implicit in the statistical model. We also focus on density-free generative AI methods. There's a large literature on density-based generative methods such as normalized flows or diffusion-based methods. \cite{wang2022a,wang2022adversarial} use ABC methods and classification to solve posterior inference problem. 

The idea of generative methods is straightforward. Let $y$ denote data and $\theta $ a vector of parameters including any hidden states (a.k.a. latent variables) $z$. 
First, we generate a "look-up" table of "fake" data  $ \{ y^{(i)}, \theta^{(i)}\}_{i=1}^N$. By simulating a training dataset of outputs  and parameters allows us to use deep learning to solve for the inverse map via a supervised learning problem.  Generative methods have the advantage of being  likelihood-free.  For example, our model might be specified by a forward map $ y^{(i)} = f(\theta^{(i)}) $  rather than a traditional random draw from  a likelihood function $y^{(i)} \sim p(y^{(i)}\mid \theta^{(i)})$.
Our method works for  traditional likelihood-based models but avoids the use of MCMC. Similarly, we can handle density-free priors such as spike-and-slab priors used in model selection. 

Posterior uncertainty is solved via  the inverse non-parametric regression problem where we predict $ \theta^{(i)} $ from $ y^{(i)} $ and $ \tau^{(i)} $. Moreover, if there is a statistic $S(y)$ to perform dimension reduction with respect to the signal distribution, then we fit an architecture of the form 
$$
\theta^{(i)}  =  H(  S(y^{(i)}) , \tau^{(i)} ) .
$$
Specifying $H$ is the key to the efficiency of the approach.  \cite{polson2024generative} propose the use of quantile neural networks implemented with ReLU activation functions.

The training dataset acts as a supervised learning problem  and allow us to  represent the posterior as a map from input $y^{(i)}$ and output $ \theta^{(i)}$.
A deep neural network is an interpolator and provides an  optimal transport map from output  to an independent  base distribution $ \tau^{(i)} $.
The base distribution is typically uniform. This doesn't have to be the case, for example one could use a very large dimensional Gaussian vector. The parameters of the neural network do not need to be identified. The training dataset acts as a procedure to find an interpolator. 
The map will provide a probabilistic representation of the posterior for \emph{any} data vector.  

The question is whether our DNN will generalize well. This is an active area of research and there is a double descent phenomenon that has been found for the generalization risk, see \cite{belkin2019,bach2023}. Given an observed $ y = y_{obs} $ we simply plug-into the network. The interpolation property of deep learners is a key feature of our generative AI method as opposed to kernel-based generative methods such as approximate Bayesian Computation (ABC) which use accept-reject methods  to calculate the posterior at a given
output.  \cite{belkin2019} pointed out a fascinating empirical property of deep learners in terms of their interpolation approximation properties, see also \cite{bach2023}. There is a second bias-variance trade off in the out-of-sample prediction problem. One of the major folklore theorems of deep learners that our generative method provides a good generalisation.

To extend our generative method to  MEU problems, we  assume that the utility function $U$  is given. Then we simply  draw additional  associated utilities $U^{(i)}_d  \defeq U(d,\theta^{(i)})$ for a given decision $d$ to add to our training dataset. Again the  baseline distribution $ \tau^{(i)} $ is appended to yield a new training dataset 
\[
\{U_d^{(i)},y^{(i)}, \theta^{(i)}, \tau^{(i)}\}_{i=1}^N.
\]
Specifically, we construct a non-parametric estimator of the form 
\[
	U_d^{(i)} = H(S(y^{(i)} ),\theta^{(i)}, \tau^{(i)} , d),
\]
where $H$ is a neural network that requires to the modeler to be specified and trained using the simulated data. The function $S$ is a summary statistic 
which allows for dimension reduction in the signal space. A number of authors have discussed the optimal choice of summary statistics, $S$. For example, \cite{jiang2017,albert2022}, use deep learning to learn the optimal summary statistics. We add another layer $H$ to learn the full posterior distribution map, see also  \cite{beaumont2002,papamakarios2018,papamakarios2019,schmidt-hieber2020,gutmann2016}

 Given that the posterior quantiles of the distributional utility, denoted by $F^{-1}_{U|d,y } ( \tau )  $ are represented as a quantile neural network, we then use a key identity shows how to represent any expectation as a marginal over quantiles, namely
$$
E_{ \theta| y} \left[ U(d , \theta \right] = \int_0^1 F^{-1}_{U |d, y} (\tau) d \tau 
$$
This is derived in \ref{sec:calculation}. The optimal decision function, $d^\star(y) \defeq \arg max_d E_{ \theta| y} \left[ U(d , \theta \right]$, 
 simply maximizes the expected utility.  
This can then be approximated via Monte Carlo and optimized over any decision variables.  We show that quantiles update as composite functions (a.k.a. deep learners) and that the Bayes map can be viewed as a concentration function. The  Lorenz curve of the utility function can be used to prove the key identity above where expectations are written as marginals of quantiles. There is a similarity with  nested sampling \cite{skilling2006} and vertical-likelihood Monte Carlo \cite{polson2015}.

Our approach focuses on generative density-free quantile methods. Quantile neural networks (QNNs) implemented via deep ReLU networks have good theoretical \cite{padilla2022,bos2024,polson2018a} and practical properties,  \cite{polson2024generative}. \cite{white1992} provides standard non-parametric asymptotic bounds in $N$ for the approximation of conditional quantile functions. \cite{polson2024generative} propose the use of quantile posterior representations and the use of ReLU neural networks to perform this task. Rather than dealing directly with densities and the myriad of potential objective functions, we directly model any random variables of interest via a quantile map to a baseline uniform measure. Our  neural estimator  network directly approximates the posterior CDF and any functions of interest.  To solve maximum expected utility problems, we simply add a given utility function  as the first layer of the network architecture. 

Another class of estimators are those based on Kernel methods, such as approximate Bayesian computations (ABC). ABC methods differ in the way, that they generate their ``fake'' look-up table. Rather than providing a neural network estimator for any output $y$, ABC methods approximate the likelihood function by locally smoothing using a circle of radius $\epsilon$ around the observed data. This can be interpreted as nearest neighbor model, see \cite{polson2024generative} for a discussion.  The advantage of ABC is that the training data set it ``tilted'' towards the observed $y$, the disadvantage is that it uses accept-reject sampling that fails in high-dimensions.   \cite{schmidt-hieber2020} provides theoretical bounds for generalisability of non-parametric kernel methods. 

The rest of the paper is outlined as follows. Section \ref{sec:genai} we provide the description of the generative AI model for learning the utility function.  Section \ref{sec:dual} provide a link with the dual theory of expected utility due to \cite{yaari1987}. We introduce the Lorenz curve of the utility function and quantile methods as a way of estimating the posterior expected utility. Section \ref{sec:application} provides an application to portfolio learning. We show to use generative methods for the normal-normal learning model and to find an optimal portfolio allocation problem based on the Kelly criterion  \cite{jacquier2012a}. Section \ref{sec:discussion} concludes with directions for future research.

\subsection{Generative Bayesian Computation (GBC)}\label{sec:genai}

To fix notation. Let $ \mathcal{Y} $ denote a locally compact metric space of signals, denoted by  $y$, and $ \mathcal{B} ( \mathcal{Y} ) $ the Borel $\sigma$-algebra of $ \mathcal{Y} $.
Let $ \lambda $ be a measure on the measurable space of signals $ (  \mathcal{Y} ,  \mathcal{B} (  \mathcal{Y}  ) ) $. 
Let $  P( d y | \theta  ) $ denote 
the conditional distribution of signals given the parameters.  
Let $ \Theta $ denote a locally compact metric space of admissible parameters (a.k.a. hidden states and latent variables $ z \in \mathcal{Z} $) and $ \mathcal{B} ( \Theta ) $ the Borel $\sigma$-algebra of $ \Theta $.
Let $ \mu $ be a measure on the measurable space of parameters $ ( \Theta ,  \mathcal{B} ( \Theta ) ) $. Let $ \Pi ( d \theta | y  ) $ denote the conditional distribution of the parameters given the observed signal $y$ (a.k.a., the posterior distribution). In many cases, $ \Pi $ is absolutely continuous with density $ \pi $ such that 
$$
 \Pi ( d \theta | y  ) = \pi (\theta  | y ) \mu ( d \theta ) .
$$
Moreover, we will write $  \Pi ( d \theta ) =  \pi (\theta  ) \mu ( d \theta )  $ for prior density $ \pi $ when available.

Our framework allows for likelihood and density free models. In the case of likelihood-free models, the output is 
simply specified by a 
map (a.k.a. forward equation)  
$$
 y=f(\theta) 
 $$ 
When a likelihood $ p( y | \theta ) $ is available w.r.t. the measure $ \lambda $, we write
$$
P ( d y | \theta   ) = p( y | \theta  ) \lambda ( d y ) .
$$
There are a number of advantages of such an approach primarily the fact that they
are density free.  They use simulation methods and deep neural networks to invert the prior to posterior map.  We build on this framework and show how to incorporate utilities
into the generative procedure.

\vspace{0.1in}

\paragraph{Noise Outsourcing Theorem} If $ (Y , \Theta ) $ are random variables in a Borel space $  ( \mathcal{Y} , \Theta ) $ then there exists an r.v. $\tau \sim U(0,1)$ which is independent of $ Y$ and a function $ H : [0,1] \times \Theta \rightarrow \mathcal{Y} $ such that
$$
(Y , \Theta )  \stackrel{a.s.}{=} (Y , H(  Y , \tau ) ) 
$$
Hence the existence of $H$  follows from the noise outsourcing theorem \cite{kallenberg1997foundations,teh2019statistical}. Moreover, if there is a statistic $S(Y)$ with $ Y \independent \Theta | S(Y) $, then 
$$
\Theta   \stackrel{a.s.}{=}  H(  S(Y) , \tau ) .
$$
The role of $S(Y)$ is equivalent to the ABC literature. It performs dimension reduction in $n$ the dimensionality of the signal.  Our approach then  is to use  deep neural network first to calculate the inverse probability map (a.k.a posterior) $ \theta \stackrel{D}{=} F^{-1}_{ \theta | y } (U ) $ where $ U$ is a vector of uniforms. In the multi-parameter case, we use an RNN or autoregressive structure where we model a vector via a sequence $ ( F_{\theta_1} ( \tau_1 ) , F_{ \theta_2 | \theta_1} ( \tau_2 ) \, \ldots ) $,

As a default choice of network architecture, we will use a ReLU network for the posterior quantile map. The first layer of the network is given by the utility function and hence this is what makes the method different from learning the posterior and then directly using naive Monte  Carlo to estimate expected utility. This would be inefficient as quite often the utility function places high weight on region of low-posterior probability representing tail risk.

\vspace{0.1in}

\paragraph{Bayes Rule for Quantiles} 
\cite{parzen2004} showed that quantile models are direct alternatives to other Bayes computations. Specifically, given $F(y)$, a non-decreasing and continuous from right function.  We define 
$$Q_{\theta| y} (u) \defeq  F^{-1}_{\theta|y}  ( u ) = \inf \left ( y : F_{\theta|y} (y) \geq u \right ) $$ which is non-decreasing, continuous from left.
\cite{parzen2004}   shows the important  probabilistic property of quantiles
$$
\theta \stackrel{P}{=} Q_\theta ( F_\theta (\theta ) ) 
$$
Hence, we can increase the efficiency by ordering the samples of $\theta$ and the baseline distribution as the mapping being the inverse CDF is monotonic. 

Let  $ g(y)$ to be a non-decreasing and continuous from left  with 
$g^{-1} (z ) = \sup \left ( y : g(y ) \leq z \right ) $.
Then,  the transformed quantile has a compositional nature, namely 
$$
Q_{ g(Y) } ( u ) = g ( Q (u ))
$$
Hence, quantiles act as  superposition (a.k.a. deep Learner). 

This is best illustrated in the Bayes learning model.  We have the following result updating prior to posterior quantiles known as the conditional quantile representation
$$
Q_{ \theta | Y=y } ( u ) = Q_\theta ( s )  \; \; 
{\rm where} \; \;   s = Q_{ F(\theta) |  Y=y } ( u) 
$$  
To compute $s$, by definition
$$
u = F_{ F(\theta ) | Y=y} ( s  ) = P( F (\theta ) \leq s | Y=y ) 
= P( \theta \leq Q_\theta (s ) | Y=y )  = F_{ \theta | Y=y } ( Q_\theta ( s ) ) 
$$

\paragraph{Maximum Expected Utility} Decision problems are characterized by a utility function $ U( \theta , d ) $ defined over parameters, $ \theta $,  and decisions, $ d \in \mathcal{D} $. 
We will find it  useful to define the family of utility random variables indexed by decisions defined by 
$$
U_d \defeq U( \theta , d ) \; \; {\rm where} \; \; \theta \sim \Pi ( d  \theta ) 
$$
Optimal Bayesian decisions \cite{degroot2005optimal} are then defined by the solution to  the  prior expected utility
\[
U(d) = E_{\theta}(U(d,\theta)) = \int U(d,\theta)p(\theta)d\theta,
\]
$$
d^\star = {\rm arg} \max_d U(d) 
$$
When information in the form of signals $y$ is available, we need to calculate the posterior distribution $p( \theta | y ) = f(y | \theta ) p( \theta ) p(y)$. Then we have to  solve for the optimal  \emph{a posterior} decision rule $ d^\star (y) $ defined by 
$$
d^\star(y)  = {\rm arg} \max_d  \; \int U( \theta , d ) p( \theta | y ) d \theta 
$$
where expectations are now taken w.r.t. $ p( \theta | y) $  the posterior distribution. 

\section{Generative Expected Utility}

Generative AI require only the ability to simulate from all distributions under consideration, signals and parameters.  
Furthermore, we can construct the posterior  $\theta \sim p( \theta | y )$.
In a stylized parametric model, we have a joint density $ p( y , \theta ) = p( y | \theta ) \pi( \theta ) $. 

The posterior distribution is given by
$$ p( \theta | y ) = p( y | \theta ) \pi( \theta ) /p(y) $$ 
where $ p(y) $ is the marginal distribution of the data. 

This induces a distribution of utilities defined for the family of r.v.s $ U_d \stackrel{D}{=} U(d, \theta )  $ where $ \theta \sim p( \theta | y ) $. This nonlinear map is then estimated using the  training data-set of utility simulations.

Imagine that we have a look-up table of variables 
 \begin{table}[H]
 \centering
 \begin{tabular}{ll}
 $y =$&		outcome  of  interest\\
 $\theta =$&	parameters\\
 $d= $&		decision variables\\
 $\tau =$&	        baseline  variables
 \end{tabular}
 \end{table}
Decision problems under uncertainty are characterized by a utility function $ U(  d , y , \theta  ) $ defined over decisions, $ d \in \mathcal{D} $, signals $ y \in \mathcal{Y} $ and
parameters, $ \theta \in \Theta $.  The \emph{a priori} expected utility is defined by \cite{degroot2005optimal} as 
$$
u(d) = E_{y,\theta}(U(d,y,\theta)) = \int U(d,y,\theta)d \Pi ( y , \theta ) .
$$

The \emph{a posteriori} expected utility for decision function, $d(y)$, is given by
$$
u(d,y) = E_{\theta|y}(U(d,y,\theta)) = \int U(d,y,\theta) d F_{\theta | y} ( \theta ) 
$$
with  expectation taken w.r.t to posterior cdf. 

The distributional form is found by defining the family of utility random variables indexed by decisions defined by 
$$
U_{d,y} \stackrel{D}{=} U(d , y  , \theta  ) \; \; {\rm where} \; \; \theta \sim \Pi (d y ,  d  \theta ) 
$$
Then we write
$$
u(d,y) = E_{ U \sim U_{d,y}  } ( U ) 
$$
This makes clear the fact that we can view the utility as a random variable defined as a mapping (a.k.a. optimal transport) of $ (y,\theta) $ evaluated at $d$.
Now we need
$$
d^\star (y) = {\rm arg} \max_d u(d,y)  .
$$
Our deep neural estimator then takes the form 
\[
U_{d,y} \stackrel{D}{=} U(d, H(S(y),\tau)).
\]
As $\theta \sim \pi(\theta)$ and $d$ is fixed, we define the utility random variable $U_d = U(d,\theta)$. Generative AI will model $\theta$ as a mapping from the data $y$ and the quantile $\tau$ as a deep learner. The nonlinear map is then estimated using simulates training data-set of utilities, signals and parameters denoted by the set  $ \{ U^{(i)} , y^{(i)} , \theta^{(i)} \} $.
We augment this training dataset with a set of independent baseline variables $ \tau^{(i)} , 1 \leq i \leq N $.

\paragraph{Latent States.}  We allow for  the possibility of further hidden states $z$ in the parameter.Our method clearly extends the models that also have hidden states (deterministic or stochastic) for example many econometric models have deterministic models for the states (e.g. DGSE models). Hence, our methods are particularly useful for dynamic learning  in economics and finance, where other methods, such as MCMC are computationally prohibitive. We illustrate our method with a simple example of a normal-normal model and portfolio allocation problem. Another class of models where our methods are particularly efficient, where there is structured sufficient statistics (that can depend on hidden latent states) that naturally performs dimensionality reduction for posterior parameter learning, see \cite{smith1992,lopes2012a}.

\subsection{Calculating  Expected Utility}\label{sec:calculation}

Expected utility is estimated using a quantile re-ordering trick and then the optimal decision function maximizes the resulting quantity. We propose using a quantile neural network as the nonlinear map. Notice, that we assume that the training data is simulated by the model that is easy to sample and the simulation costs are low. We can make $N$ as large as we want. The key to generative methods is that we directly model the random variable $\theta$ as a non-linear map (deep learner) from the data $y$ and the quantile $\tau$. This is a generalization of the quantile regression to the Bayesian setting.

\paragraph{Quantile Re-ordering} \cite{dabney2017} use quantile neural networks for decision-making and apply quantile neural networks to the problem of reinforcement learning. Specifically, they rely on the fact that expectations are quantile integrals. Let $F_U(u)$ be the CDF of the distributed utility The key identity in this context is the Lorenz curve
$$
E_{U \sim U(d , \theta ) } (U) = \int_0^1 F_U^{-1} ( \tau ) d \tau .
$$
This key identity follows from the identity 
$$
 \int_{- \infty}^\infty u d F_U(u) = \int_0^1 F_U^{-1} ( \tau ) d \tau 
$$
which holds true under  the simple transformation $\tau = F_U(u)$, with Jacobian $d\tau = f_U(u)du$.

\paragraph{Utility Lorenz Curve} The quantile identity also follows from the Lorenz curve of the utility r.v. as follows. We can compute $ \mathbb{E}(U)$ using the mean identity for a positive random variable and its
CDF or equivalently, via the Lorenz curve 
\begin{align*}
 E(U) & = \int_0^\infty (1 - F_U(u) )d u = \int_0^\infty S(u)du  \\
E(U) &= \int_0^1 F_U^{-1}(s) ds =  \int_0^1 \Lambda (1- s ) ds = \int_0^1 \Lambda(s) ds  
\end{align*}
We do not have to assume that $ F^{-1_U}(s)$, or equivalently $ \Lambda (s) $, is available
in closed form, rather we can find an unbiased estimate of this by simulating the Lorenz curve.

The Lorenz curve, $ \mathcal{L} $ of $U$ is defined in terms
of its CDF, $F_U(u)$, as 
\begin{align*}
  \mathcal{L}(u) & = \frac{1}{Z} \int_0^u F_U^{-1} ( s ) d s \; \; \text{ where } \; u \in [0,1] \\
   \mathbb{E}_{ U \sim U(d , \theta ) }  ( U)  & = \int_{ \mathcal{\Theta} } U(d,\theta) \Pi(d \theta ) \; .
\end{align*}
One feature of a Lorenz curve is that is provides a way to evaluate 
$$ \mathbb{E}(U) = \int_0^1 F_U^{-1}(s) d s $$
Hence, we only need to approximate the quantile function $  F_U^{-1}(s) $ with a deep Bayes neural estimator.

\subsection{GenBayes-MEU Algorithm}\label{sec:algorithm}

The method will generalize to the problems of the form
\[
	\arg\max_d  u(d,y) = \int U(\theta,d)p(\theta\mid d,y)d\theta
\]
First, rewrite the expected utility in terms of posterior CDF of a random variable $U_d = U(d,\theta)$, where $\theta\sim p(\theta\mid d,y)$ and $U_d$ is simply a 
transformation of $F^{-1}_{U_{d,y}}(z)$, approximated by a quantile neural network (QNN). 
We will further approximate the approximate CDF with a quantile neural network. This is a function approximation, which can be achieved using deep learning. 

Given the deep learner
\[
U_d = U(H(S(y),\tau),d)
\]
as a function of base distribution $\tau$ and the data $y$, we use $y_{obs}$ to draw value of $U$ from $\tau$. Then we can use Monte Carlo to estimate the expected utility
\[
\hat U^* = \frac{1}{N} \sum_{i=1}^N U_d^{(i)}.
\]
The algorithm starts by simulating forward $\{y_i,\theta_i\}_{i=1}^N$ and then fitting a quantile NN to the data, which approximates the CDF inverse. 

\begin{algorithm}[H]
	\caption{Gen-AI for MEU}\label{algorithm_a}
 \begin{algorithmic}[Gen-AI-Bayes]
	\STATE Simulate $ ( y^{(i)} ,  \theta^{(i)} )_{1 \leq i \leq N}  \sim p(y\mid \theta)$ or $ y^{(i)} = f ( \theta^{(i)} )  $ and  
	$\theta^{(i)} \sim \pi(\theta)$.		
	\STATE Simulate the utility $u^{(i)} = U(d^{(i)},y^{(i)}, \theta^{(i)})$
	\STATE Train $H$ using the simulated dataset for $ i = 1 , \ldots N $, via $\hat \theta^{(i)} = H( y^{(i)},\tau^{(i)})$
	\STATE Train $U$ using the simulated dataset  $U_d = U(H(S(y^{(i)}),\tau)^{(i)},d)$  for $ i = 1 , \ldots N $
	\STATE  Pick a decision $d$ that maximizes the expected utility. We use Monte Carlo to estimate the expected utility.
	\[
		E(U_d) = \sum_{i=1}^{N}F^{-1}_{U_{d}}(u_i) \rightarrow \underset{\omega}{\mathrm{maximize}}
	\]
 \end{algorithmic}
 \end{algorithm}

To find  the $\arg\max$, we can use several approaches, including Robbins-Monro or TD learning.

A related problem is that of  reinforcement learning and the invariance of the contraction property of Bellman operator under quantile projections \citep{dabney2018}.

\section{Dual Theory of Expected Utility}\label{sec:dual}

Similar approaches that rely on the dual theory of expected utility due to \cite{yaari1987}.
How do I evaluate the risky gamble? One way to introduce a utility function on payouts and not change the probabilities and calculate $E(u(x))$ or you can have a distortion measure on the probabilities a.k.a. the survival function and leave the payouts alone and calculate the expectation of the distorted survival function. Yaari showed that you can pick distortion $G$ to be $u^{-1}$. 

Risky prospects are evaluated by a cardinal numerical scale which resembles an expected utility, expect that the roles of payments and probabilities are reversed
Under expected utility we assess  gambles according to 
\[
	E(u(X)) = \int_{0}^{\infty}u(x)p_X(x)dx  = \int_{0}^{\infty}u(x) d F_X(x)
\]
The dual theory then will order gambles according to 
\[
 \tilde{E}(u(X)) =\int_{0}^{1}g\left(1-F_X(\tau)\right)d\tau = \int_{0}^{1}g\left(S_X(\tau)\right)d\tau
\]
In many cases \citep{dabney2018} we can then simply use
\[
	\int_{0}^{1}u^{-1}\left(1-F_x(x)\right)dx
\]
\cite{yaari1987} shows that one can take $ g = u^{-1}$ and still get the same stochastic ordering of gambles.
Specifically, let $Y = u(x)$, then picking $g(u) = S_x\left(u^{-1}\left(S_x^{-1}(u)\right)\right)$ yields $S_y(t) = g(S_x(t))$ as required. Hence, the expected utility decomposes as 
\[
	E(u(X)) = \int_{0}^{\infty}S_y(t)dt = \int_{0}^{\infty}g(S_x(\tau))d\tau
\]
The function $g$ is known as a distortion function.  Its is related to the notion of a concentration function \cite{fortini1994}  \cite{kruglov1992,fortini1994,fortini1995}. Another  key insight is that $g$ can be estimated using a deep quantile NN.

\paragraph{Distortion (a.k.a. Transformation) Duality} The dual theory has the property that utility is linear in wealth (in the usual framework the agent would be risk neutral). To compensate the agent has to apply a non-linear transformation known as a distortion measure  to the probabilities of payouts. This "tilting" of probabilities is also apparently in derivatives pricing using \cite{girsanov1960} change of measure. In the dual theory we are interested in the inverses of distribution function.  $g$ is a distortion measure, but it can also be interpreted as a concentration function \cite{fortini1995}.

The dual theory is motivated by the two representations of the expected value of an r.v., namely
$$
E(X) = \int_0^\infty (1 - F_X( x) ) d x ) = \int_0^1 F^{-1}_X ( x ) d x .
$$
and
$$
E(X) = \int_0^1 F_X^{-1} ( s ) d s .
$$
We will show that the latter is more useful from a computational perspective.
Adding risky choice then transforms the inner payouts (standard expected utility) or the probabilities (dual theory).

There is also open the question of how to calculate and optimize expected utility efficiently using generative methods. We propose the use of a deep neural Bayes estimator.

Let the random utility $ U \stackrel{D}{=} u(X) $ where $ X \sim F_X  $. Let $ F_U ( u) $ be the corresponding cdf. Then we can write expected utility  as
\[
	E(U) = \int_{0}^{1} UdF_U(u) = \int_{0}^{1}(1-F_U(u))du = \int_{0}^{1}S_U(t)dt
\]
where the de-cumulative distribution (a.k.a survival)  function  $S_U(\cdot) $  is defined as
\[
S_U(t) = \mathbb{P}(U> t)  .
\]
The survival function is a non-increasing function of $t$ and $S_U(0)=0$. 

To obtain the dual theory by transforming these survival probabilities  -- native that the dual theory is linear in payouts. The distortion comes from these
"risk neural" probability. Specifically, 
\[
	EU(X) = \int_{0}^{1} g \left(S_X(t)\right)dt
\]
If $g$ is differentiable, then we get the so-called Silver formula
\[
	EU(X)  = \int_{0}^{1}t g^\prime S_X(t))dF_x(t)  \; \; {\rm with} \; \;  \int_0^1 g^\prime (S_X(t))dF_X(t) = 1.
\]
Hence, the weights can be interpreted as a tilted probability measure in the dual sense.

If $g$ is convex, then $g^\prime $ is non-decreasing and 
\[
E U(X)= \int_{0}^{1}t g^\prime S_X(t))dF_X(t) = \int_{0}^{1}\phi(t)dF_X(t) = \int_{0}^{1}\phi(F_X^{-1}(\tau))d \tau.
\]
This is linear utility function and $g^\prime (S_X(t))$ is distortion. We can write 
\[
E(U) = \int_{0}^{1}\phi(t)d(f\circ F_X)(t) = \int_{0}^{1} f(F_X(t))d\phi(t).
\]

\section{Application}\label{sec:application}

\subsection{Normal-Normal Bayes  Learning: Wang Distortion}
For the purpose of illustration, we consider the normal-normal learning model.  We will develop the necessary quantile theory to show how to calculate posteriors and expected utility without resorting to densities.  Also, we show a relationship with Wang's risk distortion measure as the deep learning that needs to be learned. 

Specifically, we observe the data $ y = ( y_1,\ldots,y_n)$ from the following model
\begin{align*}
	y_1 , \ldots , y_n  \mid \theta & \sim N(\theta, \sigma^2)\\
	\theta & \sim N(\mu,\alpha^2)
\end{align*}
Hence, the summary (sufficient) statistic $S(y) = \bar y$. A remarkable result due to \cite{brillinger2012}, shows that We can learn $S$ independent of $H$ simply via OLS. 

Given observed samples $y = (y_1,\ldots,y_n)$, the posterior is then  $\theta \mid y \sim N(\mu_*, \sigma_*^2)$ with 
$$
\mu_* = (\sigma^2 \mu + \alpha^2s) / t, \quad \sigma^2_* = \alpha^2 \sigma^2 / t,
$$
where  
$$t =  \sigma^2 + n\alpha^2 \; \; {\rm and} \; \; s(y) = \sum_{i=1}^{n}y_i
$$
The posterior and prior CDFs are then related via the
$$
1-\Phi(\theta, \mu_*,\sigma_*) = g(1 - \Phi(\theta, \mu, \alpha^2)),
$$
where $\Phi$ is the normal distribution function. Here the Wang distortion function  can be viewed as a concentration function and is defined by 
\[
g(p) = \Phi\left(\lambda_1 \Phi^{-1}(p) + \lambda\right),
\]
where
$$
\lambda_1 = \dfrac{\alpha}{\sigma_*} \; \; {\rm and} \; \; 
\lambda = \alpha\lambda_1(s-n\mu)/t .
$$
The proof is relatively simple and is as follows 
\begin{align*}
	g(1 - \Phi(\theta, \mu, \alpha^2)) & = g(\Phi(-\theta, \mu, \alpha^2)) = g\left(\Phi\left(-\dfrac{\theta - \mu}{\alpha}\right)\right)\\
	& = \Phi\left(\lambda_1 \left(-\dfrac{\theta - \mu}{\alpha}\right) + \lambda\right) =  1 - \Phi\left(\dfrac{\theta - (\mu+ \alpha\lambda/\lambda_1)}{\alpha/\lambda_1}\right)
\end{align*}
Thus, the corresponding posterior updated parameters are 
$$
\sigma_* = \alpha/\lambda_1, \quad \lambda_1 = \dfrac{\alpha}{\sigma_*}
$$
and 
$$
\mu_* = \mu+ \alpha\lambda/\lambda_1, \quad \lambda = \dfrac{\lambda_1(\mu_* - \mu)}{\alpha} = \alpha\lambda_1(s-n\mu)/t .
$$
We now provide an empirical example. 

\subsubsection*{Numerical Example}
Consider the normal-normal model with Prior $\theta \sim N(0,5)$ and likelihood $y \sim N(3,10)$. We generate $n=100$ samples from the likelihood and calculate the posterior distribution.

\begin{figure}[H]
\centering
\begin{tabular}{ccc}
\includegraphics[width=0.33\linewidth]{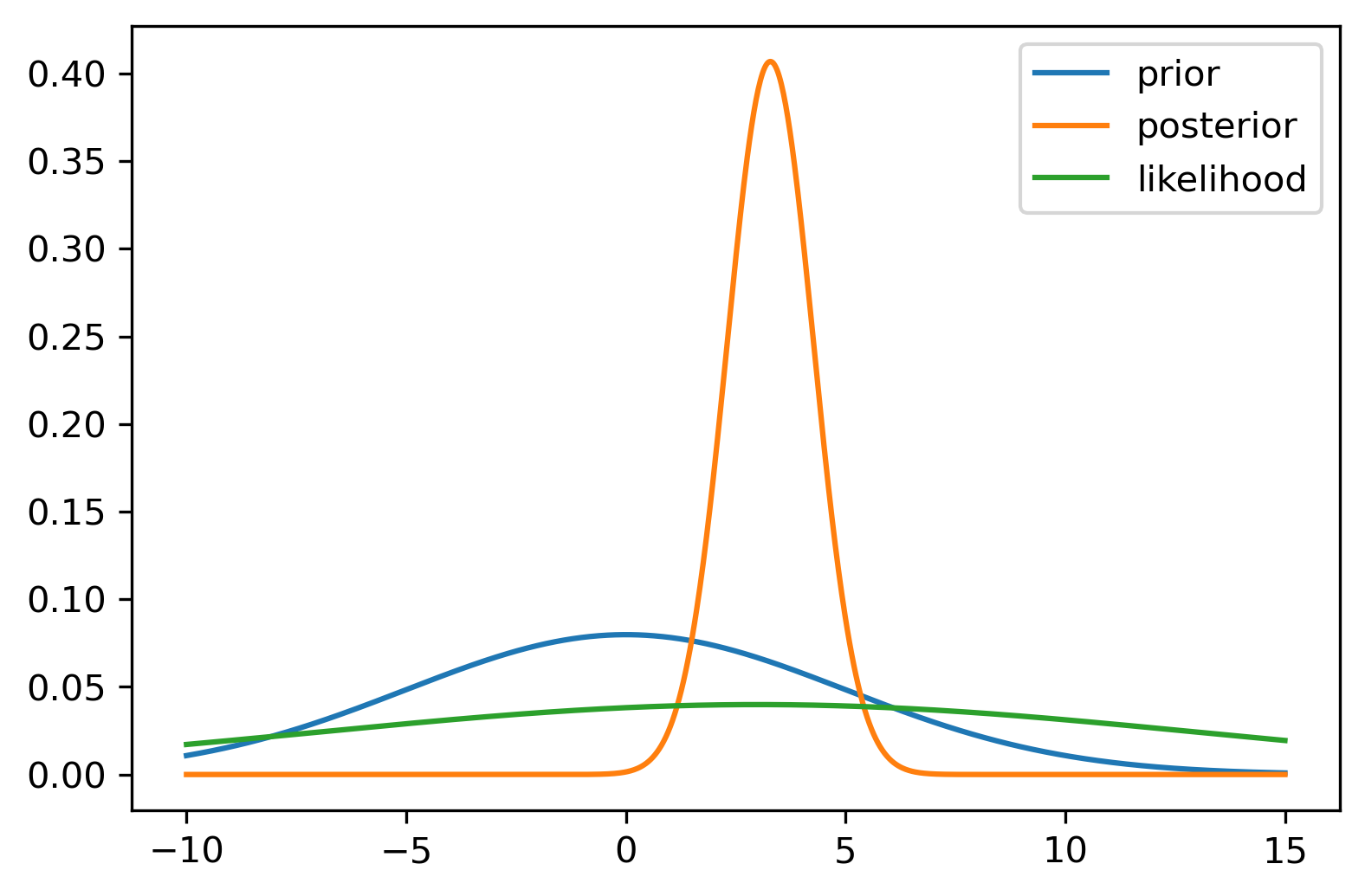} & \includegraphics[width=0.33\linewidth]{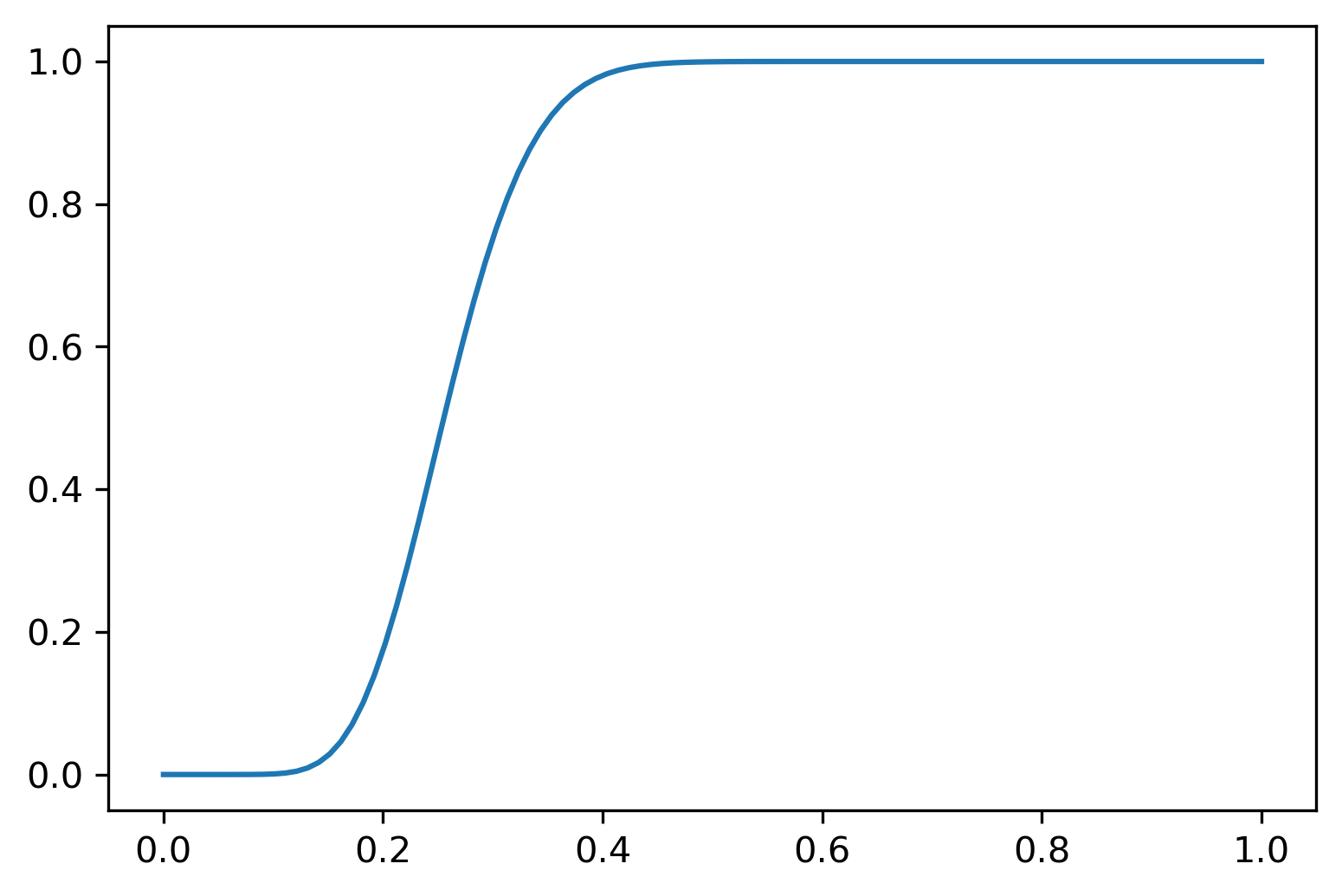} & \includegraphics[width=0.33\linewidth]{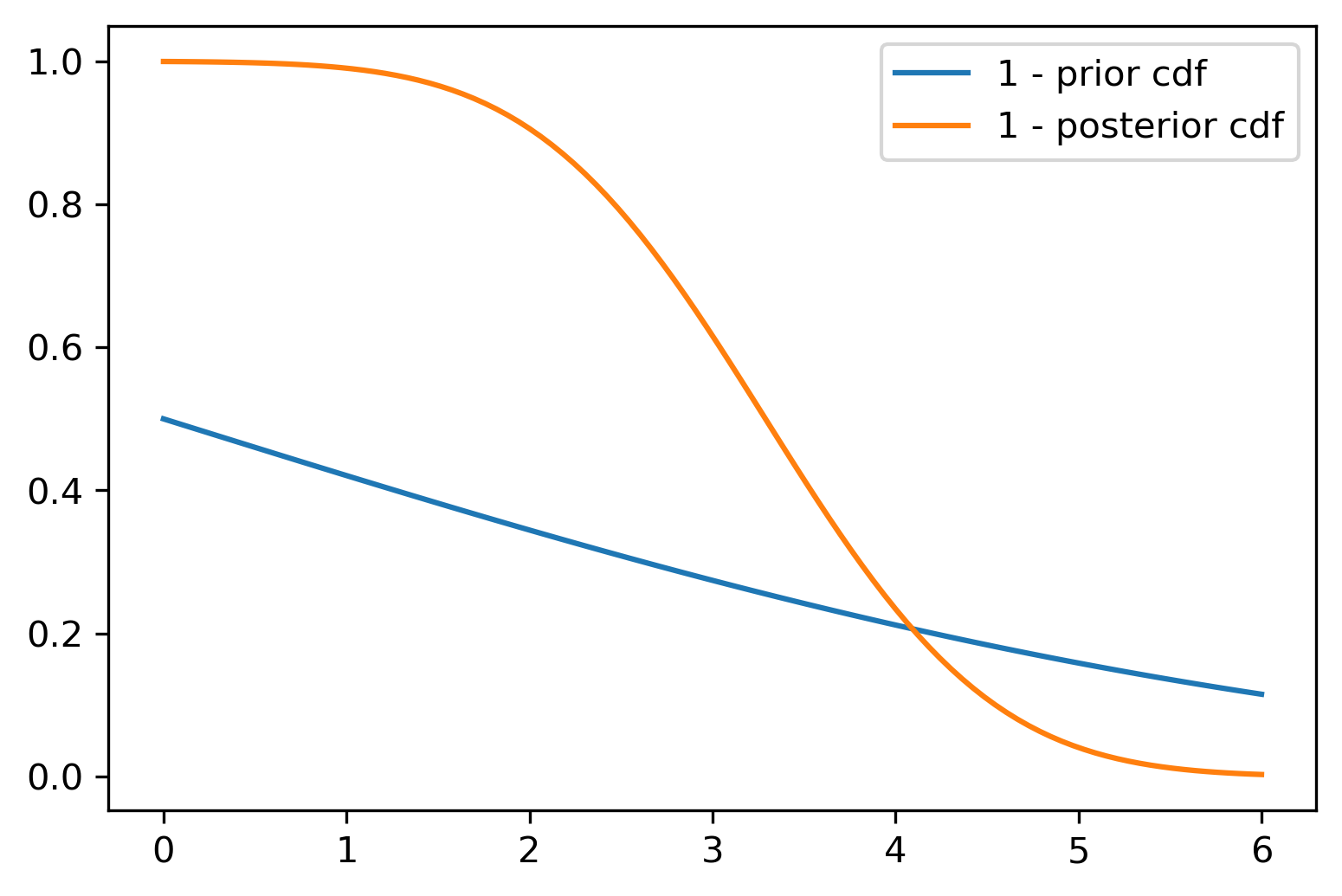}\\
(a) Model for simulated data & (b) Distortion Function $g$ & (c) 1 - $\Phi$\\
\end{tabular}
\caption{Density for prior, likelihood and posterior, distortion function and 1 - $\Phi$ for the prior and posterior of the normal-normal model.}
\label{fig:wang}
\end{figure}

The posterior distribution calculated from the sample is then $\theta \mid y \sim N(3.28, 0.98)$.

Figure \ref{fig:wang} shows the Wang distortion function for the normal-normal model. The left panel shows the model for the simulated data, while the middle panel shows the distortion function, the right panel shows the 1 - $\Phi$ for the prior and posterior of the normal-normal model.

\subsection{Portfolio Learning}

For power utility and log-normal returns (without leverage). For $\omega \in (0,1)$
\[
	U(W) = -e^{-\gamma W}, ~ W\mid \omega \sim \mathcal{N}( (1-\omega) r_f + \omega R,\sigma^2)
\]
Let $W = (1-\omega)r_f + \omega R$, with $R \sim N(\mu,\sigma^2)$, Here,  $U^{-1} $ exists and $r_f$ is the risk-free rate, $\mu$ is the mean return and $\tau^2$ is the variance of the return., Then the expected utility is 
\[
 U(\omega) = E(-e^{\gamma W}) = \exp\left\{\gamma E(W) + \frac{1}{2}\omega^2Var(W)\right\}
\]
We have closed-form utility in this case, since it is the moment-generating function of the log-normal. Within the Gen-AI framework, it is easy to add learning or uncertainty on top of $\sigma^2$ and have a joint posterior distribution $p(\mu, \sigma^2 \mid R)$. 

Thus, the closed form solution is
\[
	U(\omega) = \exp\left\{\gamma \left\{(1-\omega)r_f + \omega\mu\right\}\right\} \exp \left \{ \dfrac{1}{2}\gamma^2\omega^2\sigma^2 \right \} .
\]
The optimal Kelly-Brieman-Thorpe-Merton rule is given by 
\[
	\omega^* = (\mu - r_f)/(\sigma^2\gamma)
\]
Now we reorder the integral in terms of quantiles of the utility function. We assume utility is the random variable and re-order the sum as the expected value of $U$
\[
	E(U(W)) = \int_{0}^{1}F_{U(W)}^{-1}(\tau)d\tau
\]
Hence, if we can approximate the inverse of the CDF of $U(W)$ with a quantile NN, we can approximate the expected utility and optimize over $\omega$.

 The stochastic utility is modeled with a deep neural network, and we write
\[
	Z = U(W) \approx F, ~ W  = U^{-1}(F)
\]
Can do optimization by doing the grid for $\omega$.

The decision variable $\omega$ affects the distribution of the returns. The utility only depends on the returns $W$. Our GenAI solution is given by 
\begin{itemize}
	\item take a grid of portfolio values  $\omega_i  \in (0,1)$
	\item $W^{(i)}\mid \omega^{(i)} \sim N((1-\omega^{(i)})r_f + \omega^{(i)}\mu,\sigma^2\omega^{(i)}2)$
	\item $Z^{(i)} = U(\omega^{(i)})$, generate pairs $\left(Z^{(i)},\omega^{(i)}\right)_{i=1}^N$.
	\item Hence, $E(W) = E(Z_{\omega}) =  \int_{0}^{1}F_{Z_{\omega}}(\tau) d\tau$
	\item Learn $F_{Z_{\omega}}^{-1}$ with a quantile NN.
	\item Find the optimal portfolio weight $\omega^\star$ via 
	\[
		E(Z_{\omega}) = \sum_{i=1}^{N}F^{-1}_{Z_{\omega}}(u_i) \rightarrow \underset{\omega}{\mathrm{maximize}}
	\]
\end{itemize}

\subsubsection*{Empirical Example}
Consider $\omega \in (0,1)$, $r_f = 0.05$, $\mu=0.1$, $\sigma=0.25$, $\gamma = 2$. We have the closed-form
fractional Kelly criterion solution 
$$\omega^* = \frac{1}{\gamma}   \frac{ \mu - r_f}{ \sigma^2} = \frac{1}{2} \frac{ 0.1 - 0.05 }{ 0.25^2 } = 0.40$$ 
We can simulate the expected utility and compare with the closed-form solution.

\begin{figure}[H]
\centering
\begin{tabular}{cc}
\includegraphics[width=0.45\linewidth]{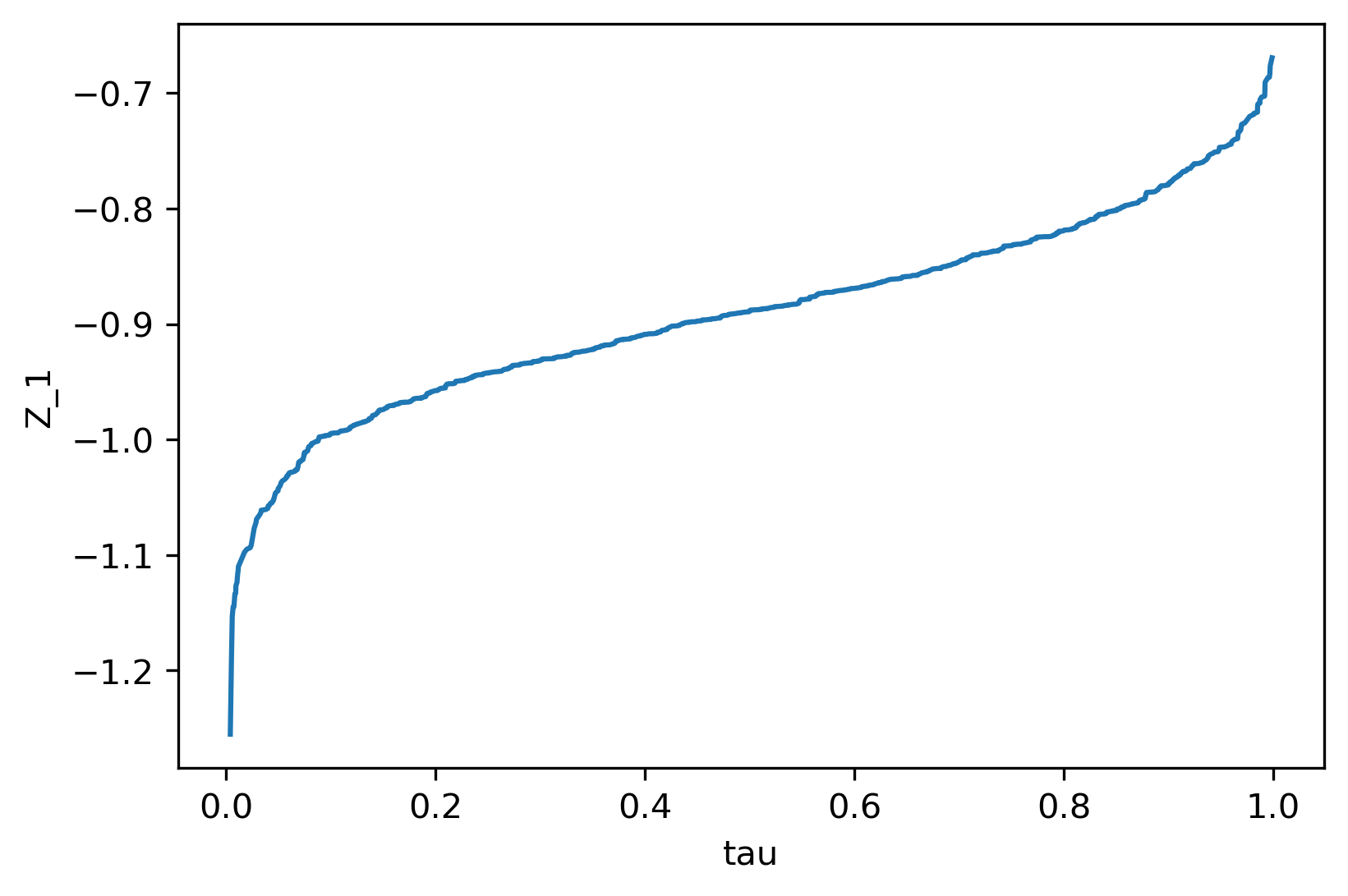} & \includegraphics[width=0.45\linewidth]{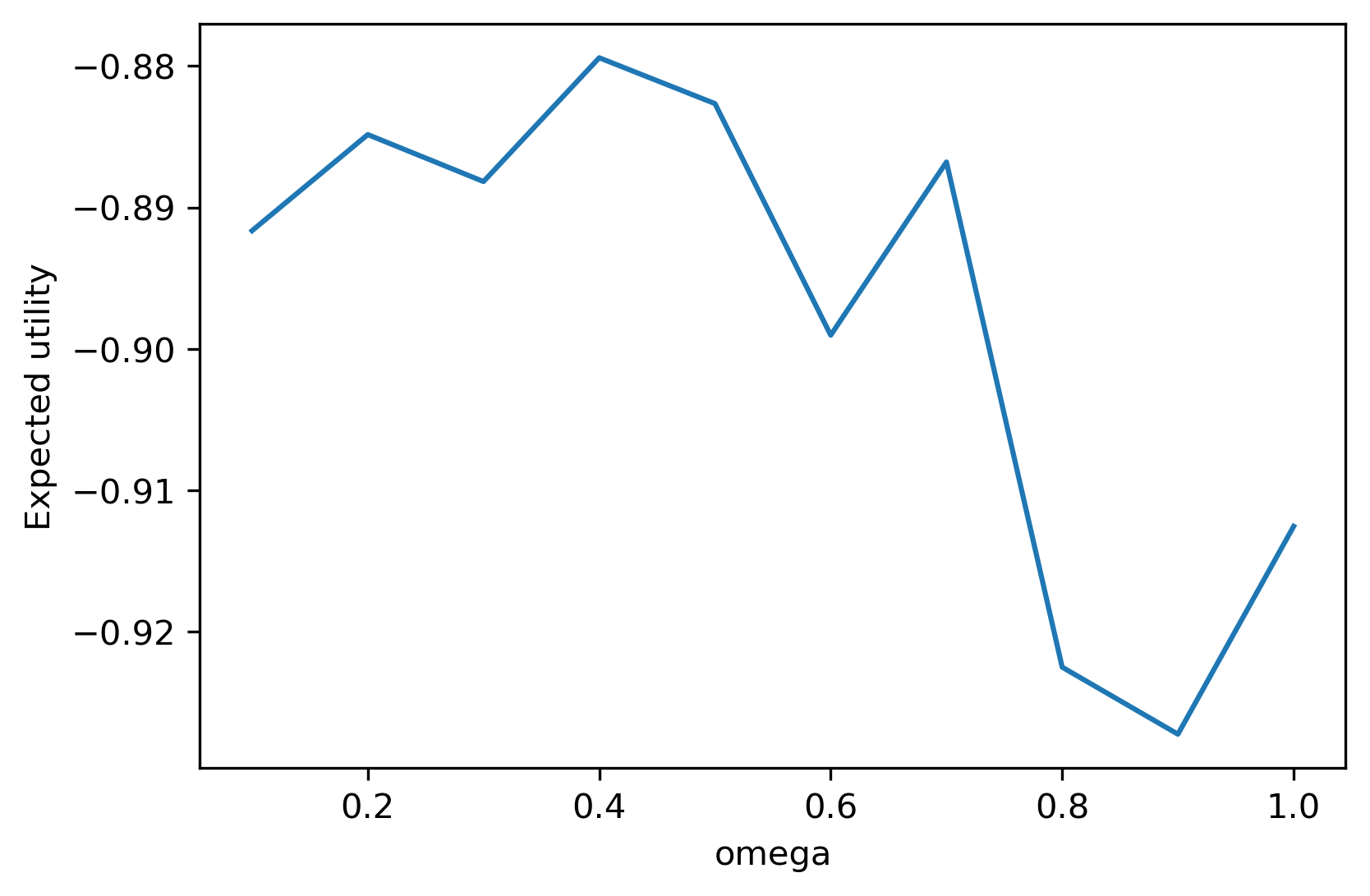}\\
\end{tabular}
\caption{Line at 0.4 optimum}
\label{fig:}
\end{figure}

\section{Discussion}\label{sec:discussion}
Generative Bayesian Computations (GBC) is a simulation-based approach to statistical and machine learning. Finding optimal decisions via maximum expected utility is challenging for a number of reasons: first, we need to calculate the posterior distribution over uncertain parameters and hidden states; second we need to perform integration to find the expected utility and third, we need to optimize the expected utility. We show how to use deep learning to solve these problems.

We propose a density-free generative method that finds posterior quantiles (and hence the posterior distribution) via deep learning estimator. Quantiles are shown to be particularly useful in solving for expected utility densities. Optimisation is then performed via a Monte Carlo approximation of the expected utility. We show how to apply this method to the normal-normal model and the portfolio learning problem. 

Our goal then was to show their use in how to  solve expected utility problems. It can be viewed as a direct implementation of  Yaari's dual theory of expected utility and to risk distortion measures that are commonplace in risk analysis.  There are many avenues for further work, for example,  the multi-parameter case and sequential decision problems are two rich ares of future research \cite{soyer2006bayesian}.

\bibliography{DistortedUtility,polsok,extra}
\end{document}